\documentclass[aps,prb,amsmath,amssymb,amstext,citeautoscript,punctuation,nofootinbib,superscriptaddress,twocolumn]{revtex4-2}
\usepackage{amsthm,mathrsfs,amsfonts,dsfont,hyperref,epsfig,braket,bm,enumerate,color,graphicx,dcolumn,charter,gensymb,comment,physics,cleveref,mathtools,natbib}
\usepackage[normalem]{ulem}
\usepackage[dvipsnames]{xcolor}
\usepackage [english]{babel}
\usepackage{graphicx}% Include figure files
\usepackage{dcolumn}% Align table columns on decimal point
\usepackage{bm}
\newcommand{\cwo}{CaWO$_4$}

\begin{document}
%\title{Phonon spectrum of CaWO$_4$ via inelastic neutron scattering and density functional perturbation theory}
\title{Coupling between CaWO$_4$ phonons and Er$^{3+}$ dopants}
\author{Mikhael T. Sayat}
\thanks{These authors contributed equally to this work.}
\address{Quantum Innovation Centre (Q.InC), Agency for Science, Technology and Research (A*STAR), 2 Fusionopolis Way, Innovis \#08-03, Singapore 138634, Singapore}
%\address{Institute of Materials Research and Engineering (IMRE), Agency for Science Technology and Research (A*STAR), 2 Fusionopolis Way, Innovis \#08-03, Singapore 138634, Republic of Singapore}
\address{Centre for Quantum Technologies, National University of Singapore, 3 Science Drive 2, Singapore 117543, Singapore}
\author{Federico Pisani}
\thanks{These authors contributed equally to this work.}
\affiliation{Institute of Physics, \'Ecole Polytechnique F\'ed\'erale de Lausanne (EPFL), CH-1015 Lausanne, Switzerland}
\author{Hin Lok Chang}
\address{National University of Singapore, Science Drive 3, Singapore 117551, Singapore}
\author{Yaroslav Zhumagulov}
\affiliation{Institute of Physics, \'Ecole Polytechnique F\'ed\'erale de Lausanne, CH-1015 Lausanne, Switzerland}
\author{Kirrily C. Rule}
\affiliation{Australian Center for Neutron Scattering, Australian Nuclear Science and Technology Organization, NSW, 2232, Australia}
\author{Tom Fennell}
\affiliation{Paul Scherrer Institut Center for Neutron and Muon Sciences, 5232 Villigen PSI, Switzerland}
%\affiliation{Laboratory for Neutron Scattering and Imaging, Paul Scherrer Institut, CH-5232 Villigen, Switzerland}
%\author{Siobhan Tobin}
%\affiliation{Australian Center for Neutron Scattering, Australian Nuclear Science and Technology Organization, NSW, 2232, Australia}
%\author{Richard Mole}
%\affiliation{Australian Center for Neutron Scattering, Australian Nuclear Science and Technology Organization, NSW, 2232, Australia}
\author{Jakob Nunnendorf}
\address{National University of Singapore, Science Drive 3, Singapore 117551, Singapore}
\author{Chee Kwan Gan}
\address{Institute of High Performance Computing (IHPC), Agency for Science, Technology and Research (A*STAR), 1 Fusionopolis Way, \#16-16 Connexis, Singapore 138632}
\author{Oleg V. Yazyev}
\affiliation{Institute of Physics, \'Ecole Polytechnique F\'ed\'erale de Lausanne, CH-1015 Lausanne, Switzerland}
\author{Ping Koy Lam}
\address{Quantum Innovation Centre (Q.InC), Agency for Science, Technology and Research (A*STAR), 2 Fusionopolis Way, Innovis \#08-03, Singapore 138634, Singapore}
%\address{Institute of Materials Research and Engineering (IMRE), Agency for Science Technology and Research (A*STAR), 2 Fusionopolis Way, Innovis \#08-03, Singapore 138634, Republic of Singapore}
\address{Centre for Quantum Technologies, National University of Singapore, 3 Science Drive 2, Singapore 117543, Singapore}
\affiliation{Department of Quantum Science and Technology, Research School of Physics, Australian National University, Canberra, ACT, 2601, Australia}
\author{Jian-Rui Soh}
\affiliation{Quantum Innovation Centre (Q.InC), Agency for Science, Technology and Research (A*STAR), 2 Fusionopolis Way, Innovis \#08-03, Singapore 138634, Singapore}
%\address{Institute of Materials Research and Engineering (IMRE), Agency for Science Technology and Research (A*STAR), 2 Fusionopolis Way, Innovis \#08-03, Singapore 138634, Republic of Singapore}
\affiliation{Centre for Quantum Technologies, National University of Singapore, 3 Science Drive 2, Singapore 117543, Singapore}
\affiliation{Department of Quantum Science and Technology, Research School of Physics, Australian National University, Canberra, ACT, 2601, Australia}
\date{\today}

\begin{abstract}
We investigate the lattice dynamics of \cwo{}, a promising host crystal for erbium-based quantum memories, using inelastic neutron scattering together with density-functional perturbation theory. The measured phonon dispersion along the (100), (001), and (101) reciprocal space direction reveals phonon bands extending up to 130 meV, with a gap between 60 and 80 meV, in good agreement with our calculations. From a symmetry analysis of the phonon eigenmodes, we identify eight Raman-active modes that can couple directly to the Er$^{3+}$ crystal-field operators, including a low-energy $B_g$ mode at 9.1\,meV that is expected to play a dominant role in phonon-assisted spin-lattice relaxation. These results provide a microscopic description of the phonon bath in \cwo{} and establish a basis for engineering phononic environments to mitigate the loss of stored quantum states and optimize Er-doped \cwo{} for quantum-memory applications.
%Rare-earth ion-doped crystals are a promising quantum memory platform for their long storage and coherence times. A promising crystal is cryogenically cooled Er$^{3+}$:\cwo{} for its atomic transitions in the C-band, integrable with existing telecommunication networks. However, lattice vibrations persist at near-zero temperatures causing decoherence to quantum states to be stored. In this work, a comprehensive mapping of the energy eigenmodes of phonons in a \cwo{} host crystal was performed. X-ray diffraction was used to map the \cwo{} host crystal in reciprocal space in accordance with a theoretically calculated map. The different energies of phonons in the Brillouin zone unit cell of \cwo{} was identified theoretically through density functional theory (DFT) and experimentally through inelastic neutron scattering using triple-axis spectrometers. The results show the existence of phonons in a low optical energy band (0--60~meV) and a high optical energy band (80--130~meV) with an approximate 20~meV band gap in between. The mitigation of these phonons could then produce ideal quantum memories with negligible decoherence leading to longer storage and coherence times with high fidelity.
\end{abstract}
\maketitle

\section{Introduction}

The realization of large-scale quantum networks relies on the development of efficient and long-lived quantum memories that enable the on-demand storage and retrieval of quantum information. Among the many candidate systems such as single atoms \cite{ritter2012elementary}, nitrogen vacancy centres \cite{yang2011high,fuchs2011quantum}, warm vapour cells \cite{pinel2013gradient,hosseini2011unconditional}, and cold atoms \cite{cho2016highly}, solid-state crystals doped with rare-earth ions have emerged as one of the most promising platforms for scalable quantum memories~\cite{thiel2011rare,zhong2015optically,ma2021one,wang2025nuclear,liu2025millisecond,ortu2022storage}. Rare-earth ion-doped crystals combine the advantages of atomic-like optical transitions with the stability and scalability of solid-state environments, enabling long storage times, high-fidelity retrieval and seamless integration with photonic architectures.

Among the available rare-earth ion dopants, erbium (Er$^{3+}$) is particularly attractive because its optical transitions ($\sim$1.5$\mu$m) lie within the C-band low-loss window of optical fibers, which enables direct compatibility and ease of integration with existing telecommunication networks \cite{heshami2016quantum}. However, Er$^{3+}$ ions possess %a total angular momentum of $J$=$15/2$, giving rise to 
a large magnetic dipole moment ($J$=$15/2$) that couples both to other magnetic species and to the lattice vibrations of the host crystal, via the spin-spin and spin-lattice interactions, respectively \cite{carnall1968electronic}. 

\begin{figure}[b!]
\includegraphics[width=0.49\textwidth]{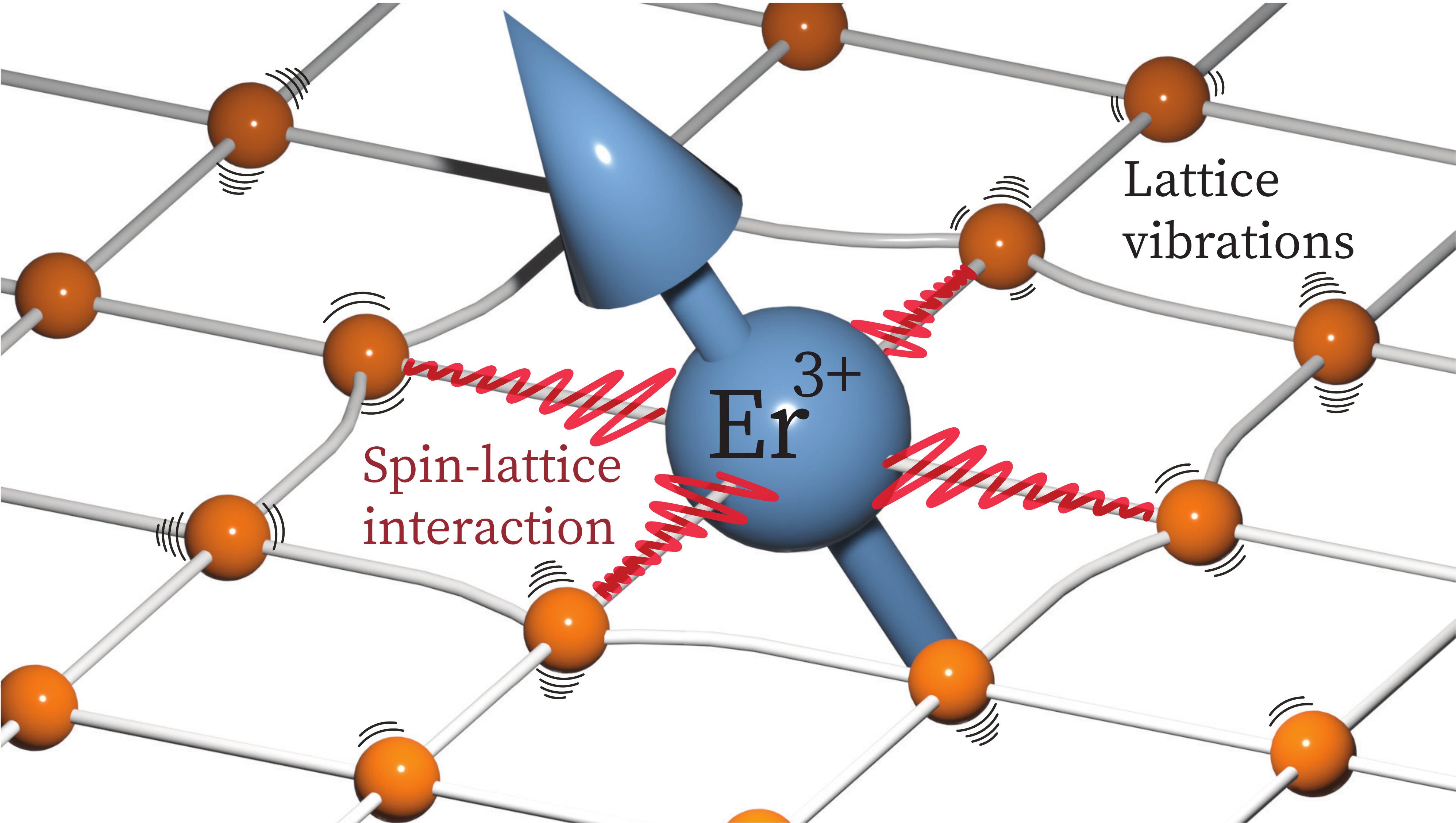}
\caption{\label{fig:Figure_1} The interaction between the large Er$^{3+}$ magnetic moment and the lattice vibrations (phonons) of the host crystal causes spin-lattice relaxation.}
\vspace{10mm}
\end{figure}

The effects of the spin-spin interaction can be mitigated by (i) selecting host crystals with exceptionally low nuclear-spin densities, such as Y$_2$SiO$_5$ and \cwo{}, (ii) operating at zero first-order Zeeman (ZEFOZ) conditions~\cite{matsuura2025explorationoptimalhyperfinetransitions,PhysRevA.85.032339,WANG2023119935} or (iii) applying strong magnetic fields~\cite{rancic_coherence_2018}. These strategies are aimed at creating a low magnetic-noise environment for the embedded Er$^{3+}$ ions, to extend their coherence times. However, while magnetic noise can be largely suppressed through these approaches, lattice vibrational noise is an intrinsic property of the crystal and remains difficult to eliminate. Phonons that couple to the large magnetic moment of Er$^{3+}$ via the spin-lattice interaction ultimately set an upper bound on how long an excited state can be maintained and therefore on the achievable coherence time.

Spin-lattice relaxation in these rare-earth ion doped systems typically occurs via three main mechanisms: the direct, Raman, and Orbach processes \cite{orbach1961spin,larson1966spin}. The Raman and Orbach mechanisms involve two- and three-phonon interactions, respectively, and their rates decrease rapidly with temperature. Hence, they can be effectively suppressed by cooling to cryogenic temperatures~\cite{Briganti_2021,budoyo2018phonon,garanin2008phonon}. In contrast, the direct process, which involves the resonant absorption or emission of a single phonon, persists even at millikelvin temperatures. These residual phonons continue to induce decoherence and ultimately lead to the loss of stored quantum information. 

A key step toward mitigating this direct-phonon induced relaxation is to identify the specific phonon modes that couple most strongly to the Er$^{3+}$ crystal-field manifold. Accomplishing this requires a detailed understanding of the complete phonon spectrum of the host crystal. However, previous studies of the lattice vibrations in \cwo{} have predominately relied on optical (infrared and Raman) spectroscopy, which are intrinsically sensitive only to zone-centre ($\Gamma$-point) phonons owing to the short probing wavelength of light~\cite{Jia2006,Goncalves2015,cavalcante2012electronic,nicol1971vibrational,BASIEV2000205,Christofilos1996pressure,Su2007Tunable, KAUR202027262,KAUR2020154804,DESOUSA2021157377}. Consequently, these techniques provide limited insight into the full phonon dispersion throughout the Brillouin zone. In principle, inelastic neutron scattering (INS) enables the measurement of phonon modes beyond the $\Gamma$ point. 

However, previous INS studies on \cwo{} by Goel \textit{et al.}~\cite{goel2014inelastic} were performed on polycrystalline samples, in which the averaging over random crystallite orientations collapses the momentum-transfer information into a one-dimensional phonon density of states (DOS), thereby losing information about the phonon dispersion across the Brillouin zone. Kesavasamy \textit{et al.}~\cite{kesavasamy1982phonon} succeeded in probing phonons away from the zone center, but their measurements were limited to energies below $\sim$37\,meV, thus excluding the high-energy optical branches. Therefore, a comprehensive experimental mapping of the phonon spectrum of \cwo{} spanning both low and high energy regimes across the full Brillouin zone is still lacking. Such measurements are essential for identifying phonon modes that couple most strongly to the Er$^{3+}$ crystal-field levels, which underpin spin–lattice relaxation and the loss of stored quantum information.

%Thus far, measurements for lattice vibrations have probed infrared and Raman active phonons using optical methods. Previous mapping probed the unit cell of \cwo up to only 37 meV~\cite{kesavasamy1982phonon}, discounting high energy phonons. In addition, the majority of measurements have been restricted to Gamma points, using powders of \cwo, due to short optical probing wavelengths \cite{gracia2011presence,cavalcante2012electronic} discounting the spatial measurement of phonons throughout the entire unit cell. This leaves a knowledge gap in the energetic measurement of phonons throughout the rest of the unit cell. 

In this work, the phonon spectra of the \cwo{} host crystal is mapped computationally, through density functional perturbation theory, and experimentally, via inelastic neutron scattering. Next, using symmetry analysis, we %The phonon spectra can then be used to 
identified the lattice vibrational modes that couple most strongly to the crystal electric field manifold of Er$^{3+}$ ions. We use this approach to identify the modes that we would target for phonon engineering to generate a phonon band gap~\cite{ma2023phonon,lutz2016modification}. This reduces lattice-induced relaxation of the stored photons and provide a pathway to develop scalable Er$^{3+}$:\cwo{} quantum memories with long storage and coherence times.

% Prior measurement of phonons of cwo using optical methods sensitive to IR and Raman active modes but  were restricted to the Gamma points due to short probing wavelength \cite{gracia2011presence,cavalcante2012electronic}. In principle, inelastic neutron scattering can probe outside the Gamam point. However \cite{GoelIndia} probed neutron scattering on powdered sample limiting it to gamma point. Need to probe outside brilluin zone, Indeed, \\cite{kesavasamy1982phonon} measured outside Gamma, but only up to 37 meV.

% \Goel: Has bandgap and similar energies to us. In results, in accordance.

%In principle, neutrons can probe outside the Brillouin zone. , while other spectroscopic work have used powders of \cwo which results in the probing of only the $\Gamma$ point and not the entire unit cell \cite{gracia2011presence,cavalcante2012electronic}.

%A method to do so is by understanding the phonon spectra of host crystals, and specifically identifying the modes that couple strongly with Er$^{3+}$. Unfortunately, the complete phonon spectra of promising host crystals such as \cwo{} have not yet been mapped.  In this work, the phonon spectra of the \cwo{} host crystal is mapped computationally, through density functional theory, and experimentally, through inelastic neutron scattering. The phonon spectra is then used to identify the phonon modes to be mitigated in order to develop Er$^{3+}$:\cwo{} quantum memories with long storage and coherence times. 

%optical papers without powder. 

\section{Methods}
%Single crystalline \cwo{} were obtained from SurfaceNet GmBH and was grown by the floating-zone method. 
Single crystals of \cwo{} grown by the floating-zone method were obtained from SurfaceNet GmBH. The quality and structure of the \cwo{} single crystals were checked with laboratory x--rays (Cu, $K_\alpha$) on a 6--circle diffractometer (Bruker). A small piece, approximately 1~mm in diameter, cleaved from the single crystal used for INS measurements, was used for this quality check. In particular, more than 200 unique reflections were measured to determine space group symmetry and crystal cell parameters of \cwo{}.%Laboratory single crystal x-ray diffraction confirmed the tetragonal $(I4_1/a)$ crystal structure of \cwo{}, 

To ascertain the lattice dynamics of \cwo{}, we performed density functional perturbation theory (DFPT) calculations of the phonon dispersion using \textsc{Quantum~Espresso}. Norm-conserving pseudopotentials~\cite{van2018pseudodojo} within the local density approximation (LDA) were employed. The wavefunctions were expanded in a plane–wave basis with a kinetic–energy cutoff of 80~Ry. The electronic Brillouin zone was sampled using a $\Gamma$–centered Monkhorst–Pack \textbf{k}–point mesh of 8$\times$8$\times$8~\cite{monkhorst1976special}. Phonon properties were computed within density–functional perturbation theory on a uniform 6$\times$6$\times$6 grid of phonon wavevectors (\textbf{q} points). The acoustic sum rule was enforced at $\Gamma$ prior to interpolation. 

In order to experimentally verify our DFPT calculations of the phonon spectrum, we performed inelastic neutron scattering measurements on the EIGER triple-axis spectrometer (TAS) at SINQ (PSI), Switzerland \cite{stuhr2017thermal} and the Taipan TAS at the Australian Centre for Neutron Scattering at the Australian Nuclear Science and Technology Organisation (ANSTO) \cite{danilkin2012taipan,rule2018recent}. The EIGER and Taipan instruments, respectively, were used to map the low (2-60 meV) and high (80-130 meV) energy phonon modes. These energy ranges were selected based on the phonon DOS measurements on \cwo{} report by Goel~\textit{et al.}~\cite{goel2014inelastic}, which found phonon in the energy ranges between 2-60 meV and 80-130 meV.

\begin{figure}[t!]
\includegraphics[width=0.49\textwidth]{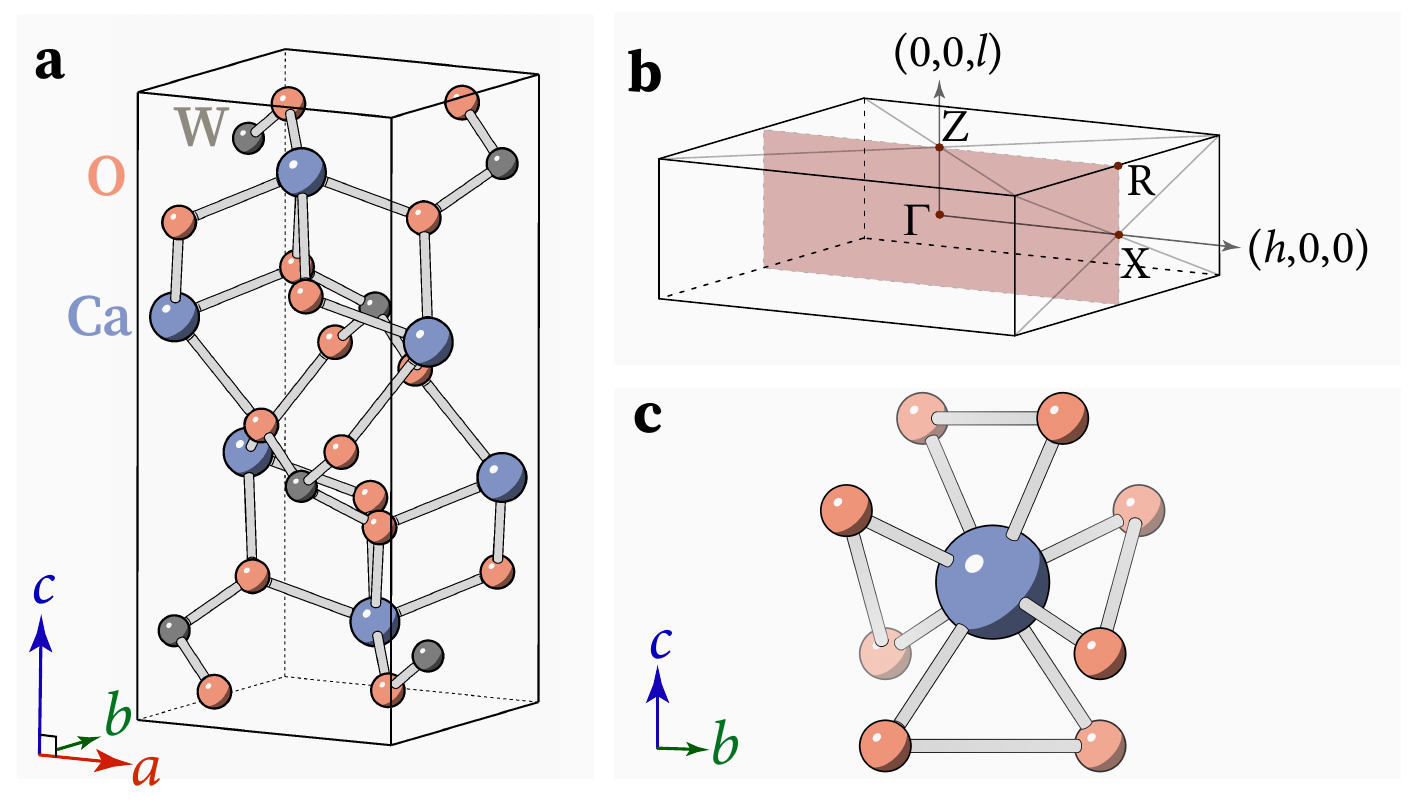}
\caption{\label{fig:Figure_2} \textbf{a} The tetragonal unit cell of \cwo{} is described by the $I4_1/a$ space group. \textbf{b} The corresponding Brillouin zone of \cwo{} with the $(h,0,l)$ scattering plane depicted in red. The labels are the high symmetry points. \textbf{c} Erbium dopants are co-ordinated by eight oxygen ligands in a ErO$_8$ polyhedron.}
\end{figure}

Phonon spectra were collected by performing energy scans, at various fixed momentum transfers, $\textit{\textbf{q}}$=$\textit{\textbf{k}}_\mathrm{f}$-$\textit{\textbf{k}}_\mathrm{i}$, along high-symmetry reciprocal space directions in the $(h,0,l)$ scattering plane [Fig.~\ref{fig:Figure_2}\textbf{b}]. These energy scans were performed by fixing the outgoing wavevector at $\textit{\textbf{k}}_\mathrm{f}$=2.66\,\AA$^{-1}$ ($E_\mathrm{f}$=14.86\,meV) and varying the incident neutron wavevector ($\textit{\textbf{k}}_\mathrm{i}$) using an incident beam monochromator to achieve the desired energy ($\Delta E$). For low energy transfers ($\Delta E$$\leq$$60$\,meV) on the EIGER instrument a pyrolitic graphite (PG) $(002)$ double-focusing monochromator was used, and for high energies ($\Delta E$$\geq$$80$\,meV) on the Taipan instrument, a Cu $(200)$ vertically focusing monochromator was used with open-40'-40'-open collimation. In both instruments, a PG $(002)$ analyzer and $^3$He tube were used to resolve and detect the scattered neutrons, respectively.

To mitigate the weak neutron cross-section of phonons, large single crystals of \cwo{} with typical dimensions of 2$\times$2$\times$1 cm$^{3}$ were used for both INS experiments. The INS experiments were conducted at $T$$=$$200$\,K. Prior to the inelastic scattering measurements, the \cwo{} single crystals were oriented on the ORION single-crystal diffractometer (SINQ, PSI). The alignment was performed such that the crystallographic $a$ and $c$ axes lay within the horizontal scattering plane, thereby allowing access to phonon excitations in the $(h,0,l)$ reciprocal-lattice plane [Fig.~\ref{fig:Figure_2}]. Because \cwo{} crystallizes in the tetragonal $I4_1/a$ space group, the $a$ and $b$ axes are symmetry equivalent. Consequently, restricting the measurements to the $a$–$c$ plane captures the essential lattice dynamics without loss of generality. 
%The INS experiments were conducted at $T$$=$$200$\,K %in order to enhance the scattering intensity from phonons by maximizing the Bose–Einstein population factor.

\begin{figure}[ht!]
\includegraphics[width=0.47\textwidth]{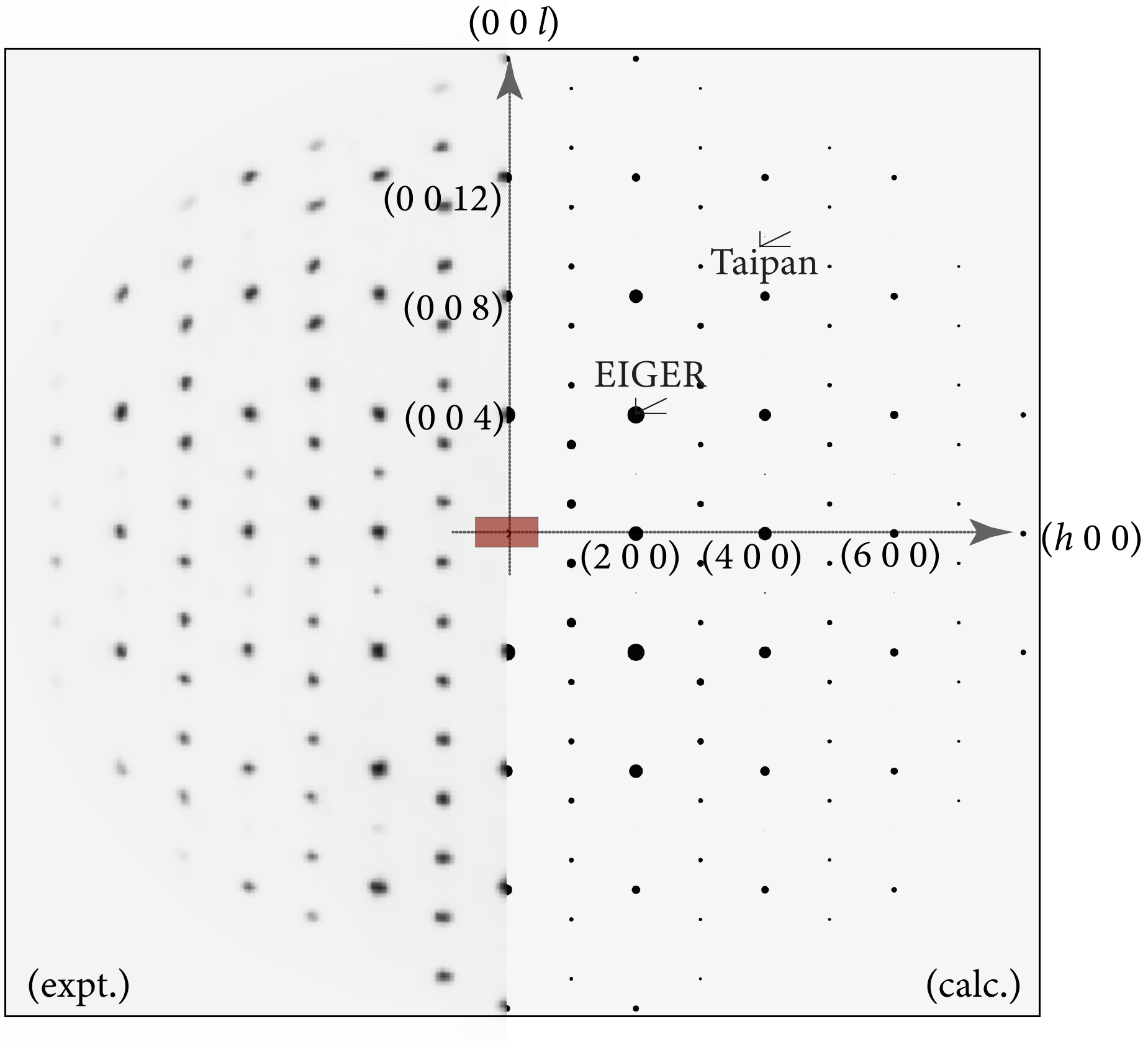}
\caption{\label{fig:Figure_3} The comparison between the experimental (left) and calculated (right) reciprocal space map of \cwo{} in the $(h,0,l)$ scattering plane obtained with single crystal x-ray diffraction. The corresponding first Brillouin zone is depicted by the red rectangle, along with the reciprocal space trajectory of the INS experiments on the EIGER and Taipan triple-axis spectrometers.}
\end{figure}

\section{Results}

\subsection{Crystal structure}

All of the observed x-ray diffraction peaks can be indexed by the tetragonal $I4_1/a$ space group. The refined lattice parameters at room temperature of $a$=$b$=5.2387(5)\,\AA{} and $c$=11.3780(10)\,\AA, are in good agreement with previously reported values~\cite{zalkin1963x,kay1964neutron,culver2015low}. The left panel of Fig.~\ref{fig:Figure_3} plots the measured x-ray diffraction peaks within the $(h,0,l)$ scattering plane. Indeed, the measured intensity distribution and pattern are in excellent agreement with the calculated reciprocal space map, shown on the right panel of Fig.~\ref{fig:Figure_3}. Along the $(0,0,l)$ direction, only reflections with $l$=$4n$ are observed, consistent with the $4_1$ screw axis along $c$, while along $(h,0,0)$ only reflections with even $h$ appear. More generally, the body-centering condition imposes $h$+$l$=$2n$ for allowed-reflections.

These extinction rules demarcate the first Brillouin zone (in red) within the $(h,0,l)$ plane, as shown in Fig.~\ref{fig:Figure_3}. The corresponding high symmetry point labels are denoted in Fig.~\ref{fig:Figure_2}\textbf{b}, with the zone center at $\Gamma$=$(0,0,0)$ and the vertical and horizontal axis corresponding to $Z$=$(0,0,\frac{1}{2})$ and $X$=$(\frac{1}{2},0,0)$, respectively. 

\subsection{Lattice Dynamics}
\begin{figure}[b!]
\includegraphics[width=0.49\textwidth]{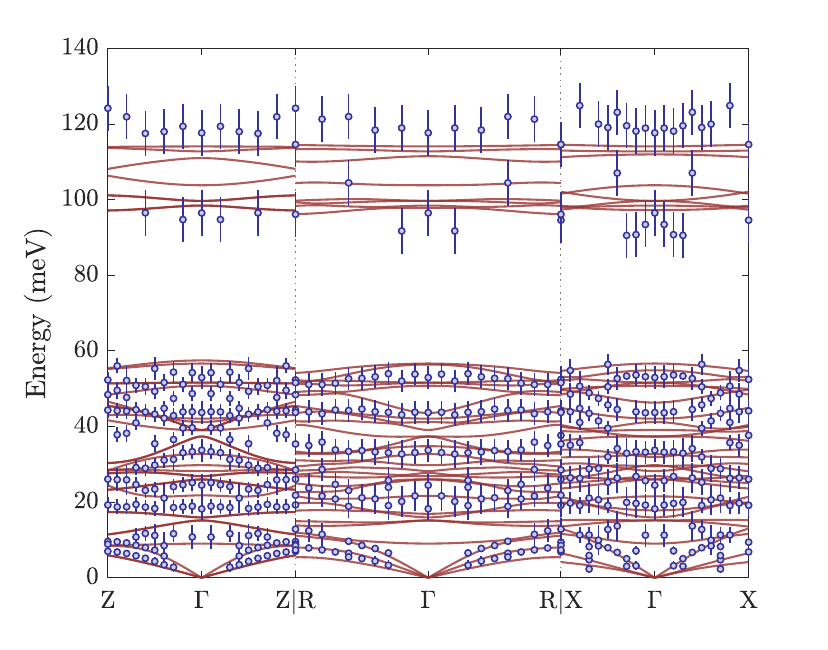}
\caption{\label{fig:Figure_4} The calculated (red lines) and experimental (blue circles) phonon dispersion of \cwo{} along the Z--$\Gamma$--Z$|$R--$\Gamma$--R$|$X--$\Gamma$--X high-symmetry path.}
\end{figure}

%The resulting phonon spectrum of CaWO$_4$, shown in Fig.~\ref{fig:Figure_4}, reflects the symmetry of the tetragonal scheelite lattice (space group $I4_1/a$, point group $C_{4h}$ at the Brillouin-zone center). Group-theoretical analysis predicts a set of Raman-active and infrared-active modes. The Raman-active sector comprises $3A_g + 5B_g + 5E_g$, while the infrared-active sector comprises $5A_u + 5E_u$. Additional silent $3B_u$ modes and the three acoustic branches complete the spectrum~\cite{}. 

%DFT predicts that the acoustic bands exist between 0 and 10~meV which corresponds to when phonons vibrate in-phase with the crystal. Low energy optical phonon bands exist between 10 and 58~meV and high energy optical phonon bands exist between 98 and 115~meV which correspond to when phonons vibrate out-of-phase with the crystal. An optical phonon band gap of approximately 40~meV resides between the low and high energy phonon bands which corresponds to the energy levels that can be occupied by Er$^{3+}$ dopants.

Having established the crystal structure of \cwo{}, we now investigate its lattice dynamics from our DFPT calculations. Figure~\ref{fig:Figure_4} plots the calculated phonon dispersion of \cwo{} (in red) following the Z--$\Gamma$--Z$\mid$R--$\Gamma$--R$\mid$X--$\Gamma$--X high-symmetry path in the $(h,0,l)$ reciprocal space plane, as defined in Fig.~\ref{fig:Figure_2}\textbf{b}. The calculations yield a total of 36 phonon branches, consistent with the 12-atom conventional tetragonal unit cell (2 formula units $\times$ 6 atoms per unit $\times$ 3 degrees of freedom).

The calculated dispersion reflects the $I4_1/a$ tetragonal symmetry of the scheelite lattice. The three acoustic modes extend up to approximately 10\,meV, followed by a broad manifold of low-energy optical modes spanning 10-58\,meV, originating predominantly from vibrations involving the heavier Ca and W atoms. A band gap of roughly 40\,meV separates these from the high-energy optical modes located between 98-115\,meV, which arise from stiff W-O bond stretching within the WO$_4$ tetrahedra~\cite{cavalcante2012electronic,YinDFT}. The presence of such a gap is a characteristic feature of scheelite-structured materials~\cite{zhang1998electronic,huang2013physical}.

To validate the DFPT calculations, the experimentally measured phonon excitations from our inelastic neutron scattering (INS) experiments are overlaid as blue markers in Fig.~\ref{fig:Figure_4}. Guided by prior INS studies on polycrystalline \cwo{}~\cite{goel2014inelastic}, we targeted both the low-energy (2-60\,meV) and high-energy (80-130\,meV) regions of the spectrum. Acoustic and low-lying optical modes were performed on the EIGER spectrometer, while the high-energy optical excitations were accessed using the Taipan spectrometer. The respective zone centers ($\Gamma$-point) are located at $(2,0,4)$ and $(4,0,10)$, to enable the mapping of phonons along the $\Gamma$-Z, $\Gamma$-X, and $\Gamma$-R directions within the $(h,0,l)$ plane (Figs.~\ref{fig:Figure_2}\textbf{b} and \ref{fig:Figure_3}).

Overall, we find a good agreement between the measured low-energy (2-60\,meV) phonon branches and calculations, confirming the reliability of DFPT in this regime. At higher energies, the measurements exhibit slightly reduced phonon energies compared to the calculated dispersion. This deviation is expected, as the DFPT calculations were performed at 0\,K, whereas the INS measurements were acquired at 200\,K, where anharmonic effects soften the phonon frequencies. Nevertheless, the measured excitations remain within the predicted spectral range and broadly follow the calculated phonon dispersion.

\begin{figure}[h!]
\includegraphics[width=0.5\textwidth]{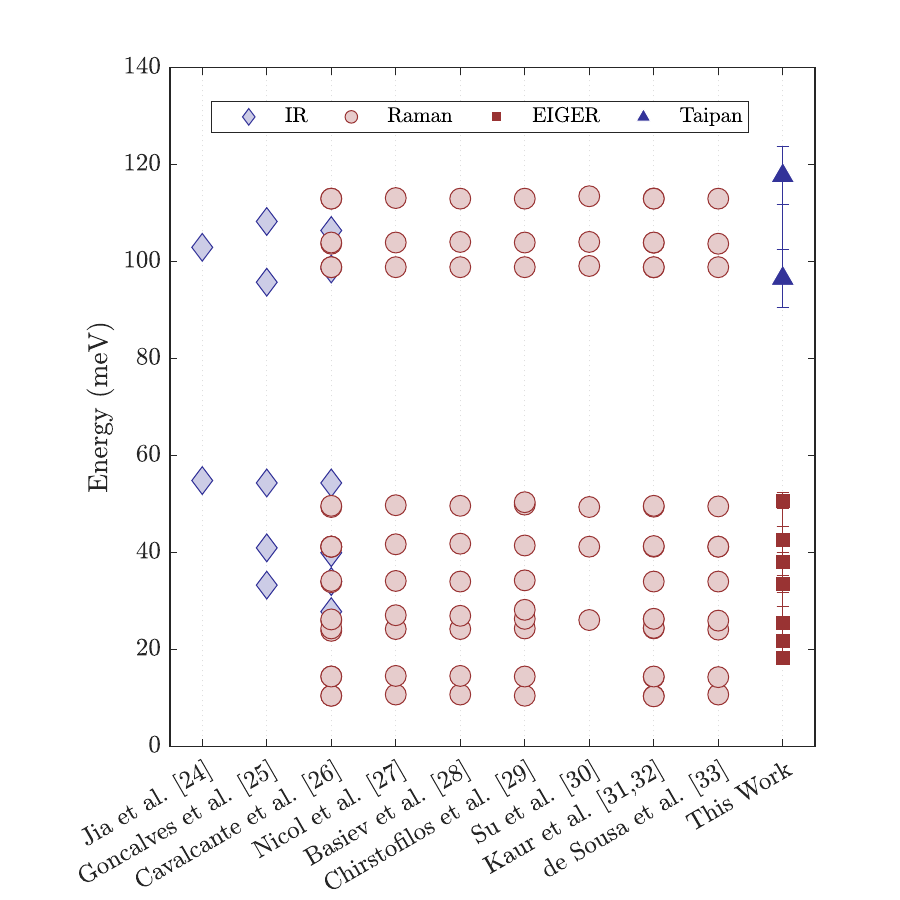}
\caption{\label{fig:GammaPoint} Comparison between previously reported infrared \cite{Jia2006,Goncalves2015,cavalcante2012electronic} and Raman \cite{cavalcante2012electronic,nicol1971vibrational,BASIEV2000205,Christofilos1996pressure,Su2007Tunable,KAUR202027262,KAUR2020154804,DESOUSA2021157377} active modes and the $\Gamma$-point phonon modes measured in this work. %Phonon energy levels at the $\Gamma$ point of infrared \cite{Jia2006,Goncalves2015,cavalcante2012electronic} and Raman active modes \cite{cavalcante2012electronic,nicol1971vibrational,BASIEV2000205,Christofilos1996pressure,Su2007Tunable, KAUR202027262,KAUR2020154804,DESOUSA2021157377} compared to this work performed on the EIGER and Taipan TAS.
}
\end{figure}

Our results are also consistent with previous spectroscopic studies. For instance, Goel \textit{et al.}~\cite{goel2014inelastic} reported a phonon density of states extending up to $\sim$130\,meV with a pronounced reduction in intensity between $\sim$60-95\,meV, consistent with the optical phonon band gap observed here. Similarly, earlier INS measurements along $\Gamma$-Z up to $\sim$37\,meV by Kesavasamy \textit{et al.}~\cite{kesavasamy1982phonon} agree well with both our DFPT predictions and experimental dispersion in that regime. Furthermore, the comparison of the measured phonon eigenmodes extracted at the $\Gamma$-point show excellent correspondence across both low- and high-energy modes, with previously reported Raman and infrared spectroscopy data [Fig.~\ref{fig:GammaPoint}].

Taken together, our INS measurements and DFPT calculations establish a robust and comprehensive phonon description of \cwo{}, and sets the stage for exploring how the lattice vibrations couple to the rare-earth Er$^{3+}$ dopants, which is  the focus of the next section.

 %addition to the energy measurements of phonons in the Brillouin zone, the phononic energies at the $\Gamma$ point was compared to previous results as shown in Fig. \ref{fig:GammaPoint}. It can be seen that both the phononic energies in the low optical regime as measured by the EIGER TAS and the phononic energies in the high optical regime as measured by the Taipan TAS are within the respective energy regimes as compared to previously measured infrared and Raman active phonons at the $\Gamma$, further validating the results presented.}

%The phonon spectrum of the \cwo{} host crystal was obtained through inelastic neutron scattering using the EIGER and Taipan instruments for the low-energy and high-energy phonon bands. }

%As shown in Fig.~\ref{fig:Figure_4}, % shows both the experimental and theoretical energies of phonons located in the Brilluoin zone. It can be seen that
%the measured phonons at low energies (2--60~meV) from EIGER are in agreement with the energies calculated using DFT. However, there is some discrepancy in the energies of measured phonons using Taipan and the DFT-calculated phonon bands at high energies (80--130~meV). This could be attributed to different temperature settings; DFT calculations were made with a temperature of 0~K whereas the \cwo{} crystal was kept at a temperature of 200~K for experimental measurements of phonons. However, it is crucial to point out that the experimental results exist in the same regime according to theory, despite this temperature difference.

%\textbf{Heatmap/9panel Figure from HinLok and explain results. The above will be adapted.} 

%\subsection{Coupling between $\Gamma$ point phonons and CEF levels of Er$^{3+}$}
\subsection{Er$^{3+}$-phonon coupling}
We now examine how the $\Gamma$-point phonons of CaWO$_4$ couple to the crystal electric field (CEF) levels of erbium. In \cwo{}, Er\(^{3+}\) occupies the Ca$^{2+}$ site with local \(S_4\) symmetry. As such only terms with $q$=0 and $q$=4 in the Stevens expansion $\hat H_{\text{CF}} = \sum_{k,q} B_k^q O_k^q $ are symmetry-allowed, which leads to the following Er\(^{3+}\) CEF Hamiltonian,
\begin{align}
 \hat H_{\text{CEF}} = B_2^0 O_2^0 + B_4^0 O_4^0 + B_4^4 O_4^4 + B_6^0 O_6^0 + B_6^4 O_6^4 .   
\end{align}

The interaction between the electronic states of Er$^{3+}$ and the phonons of CaWO$_4$ can be explained by modification of the CEF parameters by the displacement of the surrounding oxygen ligands [Fig.~\ref{fig:Figure_1}\textbf{c}]. Each phonon mode that distorts the ErO$_8$ coordination polyhedron modifies the crystal-field potential, and modulates the Stevens parameters $B_k^q$. From a group-theoretical point of view, the effect can be described as the direct product of the irreducible representation (irrep) of a phonon mode with that of a Stevens operator. Crucially, only those combinations that contain the totally symmetric representation of the site symmetry group produce non-zero matrix elements, and therefore only these phonons contribute directly to crystal-field-phonon coupling.

Formally, the dependence of the crystal-field parameters on the phonon displacement can be written as an expansion,
\begin{align}
B_k^q(\{Q_\lambda\}) &= B_k^q \,+\, \sum_\lambda \left(\frac{\partial B_k^q}{\partial Q_\lambda}\right) Q_\lambda \nonumber\\ 
&+\frac{1}{2}\sum_{\lambda\lambda'} \left(\frac{\partial^2 B_k^q}{\partial Q_\lambda \partial Q_{\lambda'}}\right) Q_\lambda Q_{\lambda'} \,+ \,\cdots ,
\end{align}
where first term corresponds to the static crystal field, while the second and third terms describe linear and quadratic crystal-field–phonon interactions, respectively. Quantization of the displacement coordinates $Q_\lambda$ in terms of creation and annihilation operators gives the following crystal-field-phonon coupling Hamiltonian
\begin{align}
    \hat H_{\text{cf-ph}}  &= \sum_{k,q,\lambda} g_{kq,\lambda}\, O_k^q\,(a_\lambda + a_\lambda^\dagger)\nonumber\\ 
&+ \sum_{k,q,\lambda\lambda'} h_{kq,\lambda\lambda'}\, O_k^q\,(a_\lambda + a_\lambda^\dagger)(a_{\lambda'} + a_{\lambda'}^\dagger)\,+\,\cdots,
\end{align}
where $g_{kq,\lambda}$ and $h_{kq,\lambda\lambda'}$ are coupling constants related to the derivatives of $B_k^q$ with respect to the phonon displacement. The linear term produces one-phonon processes, while the quadratic term allows two-phonon interactions, which can play a role in multiphonon relaxation.

The phonon eigenmodes at the Brillouin-zone center reflect the $C_{4h}$ point group symmetry of the tetragonal scheelite lattice ($I4_1/a$), and can be categorized based on their symmetry. Group-theoretical analysis predicts a set of Raman-active and infrared-active modes. The Raman-active sector comprises $3A_g$ + $5B_g$ + $5E_g$, while the infrared-active sector comprises $4A_u$ + $4E_u$, where $E$ modes are doubly-degenerate. Additional silent $3B_u$ modes and the three acoustic branches complete the spectrum. 

\begin{figure}[t!]
\includegraphics[width=0.5\textwidth]{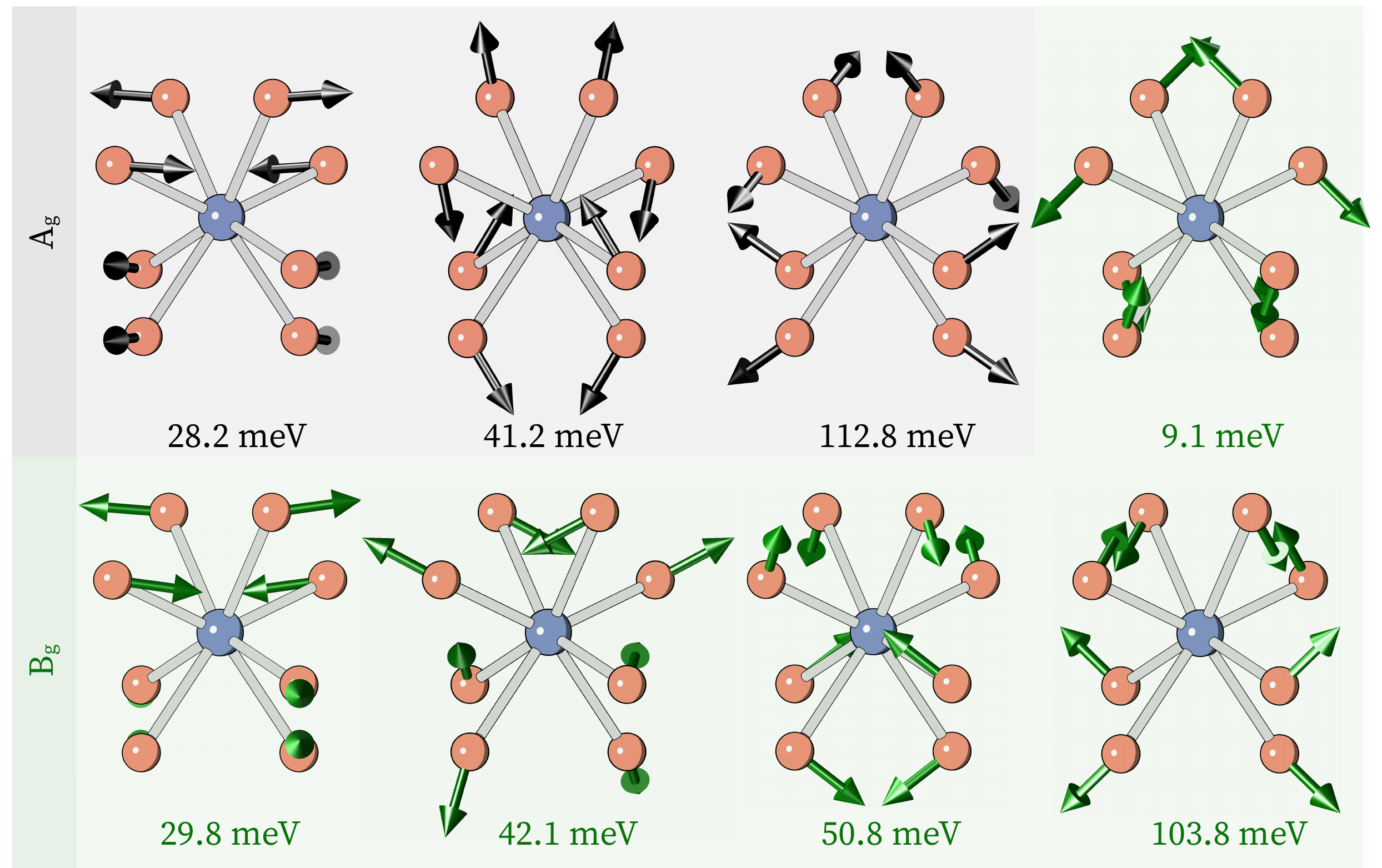}
\caption{\label{fig:Figure_6}The three $A_g$ and five $B_g$ zone-center phonon eigenmodes which can couple to the Er$^{3+}$ CEF manifold, by symmetry, and the corresponding eigen energies.}
\end{figure}

Symmetry strongly constrains which phonons participate in crystal-field-phonon scattering process. As we mentioned before, at the $S_4$ site occupied by Er$^{3+}$ in \cwo{}, the only Stevens operators that appear in the crystal-field Hamiltonian are $O_2^0$, $O_4^0$, $O_4^4$, $O_6^0$, and $O_6^4$. The corresponding phonons must have the same irrep. Raman-active $A_g$ modes couple to the axial operators $O_k^0$, while $B_g$ modes couple to the fourfold operators $O_k^4$. From our DFPT calculations, the relevant $B_g$ modes in CaWO$_4$ occur at 9.1, 29.8, 42.1, 50.8, and 103.8\,meV, respectively, and the $A_g$ modes appear at 28.2, 41.2, and 112.8\,meV. The associated 3$A_g$ and 5$B_g$ eigenmodes are plotted in Fig.~\ref{fig:Figure_6}. Other even-parity modes, such as $E_g$, do not couple linearly but can contribute through quadratic couplings. Odd-parity infrared-active $A_u$ and $E_u$ phonons are excluded from direct linear coupling by inversion symmetry. However, they can become important in higher-order processes and when inversion is locally broken by defects or strain.

\section{Discussion}
Mapping the phonon spectrum of \cwo{} experimentally, is a key step toward identifying which lattice vibrational modes are ultimately responsible for the loss of quantum information stored in rare-earth ions. Further characterisation of \cwo{}, such as its dielectric properties \cite{hartman2025dielectric}, could be performed to ascertain feasibility in different applications. Beyond quantum-memory applications, \cwo{} is also of interest as a cryogenic scintillator for rare-event and dark-matter detection, where lattice vibrations play a central role in energy relaxation and signal formation~\cite{DARKMATTER_1,DARKMATTER_2,DARKMATTER_3,DARKMATTER_4,DARKMATTER_5}. In particular, our inelastic neutron scattering results validate the DFPT-calculated phonon spectrum, which provide a basis for determining which phonon eigenmodes can couple to Er$^{3+}$ dopants, by symmetry. Such a coupling can mediate relaxation and dephasing pathways for stored quantum states, ultimately limiting both the achievable optical storage and coherence time. Even at cryogenic temperatures and in a host with a comparatively low spin bath of \cwo{}, residual phonon populations and phonon-assisted processes remain relevant. Thus, deliberate strategies to suppress, redirect, or otherwise mitigate phonon participation is required to overcome the phonon bottleneck~\cite{ma2023phonon}, rather than relying only on cryogenic cooling. One promising strategy is phononic engineering to modify the phonon dispersion of \cwo{}. 

For example, one approach is nanostructuring the host crystal to create phonon band gaps at selected eigen energies, particularly those chiefly responsible for causing the loss of stored quantum states. In \cwo{}, this includes suppressing or shifting the low-energy $B_g$ mode at 9.1 meV to higher energies, where it is less thermally populated and less effective at coupling to Er$^{3+}$ crystal-field levels. Such structures suppress the local phonon density of states, modify phonon dispersion, and reduce lattice-induced relaxation~\cite{lutz2016modification,eyring2002handbook,ma2023phonon}.

A complementary approach is to use phononic engineering to enhance heat extraction from the crystal~\cite{chia2021diamond,joe2024high}, which is especially important when substantial optical power is required to be applied during memory operation. For rare-earth–ion-doped systems, waveguides fabricated directly within the crystal can guide optical modes while allowing quantum states to propagate in regions where phonons are otherwise strongly suppressed, for example, near the natural 60–80 meV gap in \cwo{} (Fig.~\ref{fig:Figure_4}). In such engineered environments, improved heat flow helps the crystal return to its base cryogenic temperature more rapidly, enabling higher repetition rates and the use of shorter, more intense optical control pulses without overheating.

\section{Conclusion}
In conclusion, we have mapped the phonon spectrum of \cwo{} using inelastic neutron scattering and found excellent agreement with density-functional perturbation theory calculations. A symmetry analysis of the phonon eigenmodes identifies eight modes that can couple directly to Er$^{3+}$ crystal-field levels and therefore play a central role in phonon-induced spin-lattice relaxation. Finally, we outlined phonon-engineering strategies to suppress or control these modes, providing pathways toward improving the photon storage times and performance of Er-doped \cwo{} quantum memories.
%In conclusion, the phonon spectra for \cwo, a potential quantum memory host crystal candidate, was mapped theoretically, through density functional theory, and experimentally, through inelastic neutron scattering. The phonon spectra was calculated and measured as the phoninc eigenmodes within the Brillouin zone. Both experimental results using the EIGER and Taipan TAS are in good agreement with theory as well as previously measured infrared and Raman active phonons at the $\Gamma$ point -- the low and high energy regimes as well as the band gap were correctly identified. In the endeavour of developing REID Crystals such as Er$^{3+}$:\cwo as a host crystal for quantum memory applications, these phonons will need to be mitigated through phonon engineering methods. This will minimise decoherence resulting in long storage and coherence times. Conversely, the phonons can also be used to encode information. They will need to be extracted effectively through similar processes and act as quantum information carriers. 
%\newpage
\section{Data Availability}

The data that support the findings of this study are available from the corresponding author upon reasonable request.

\begin{acknowledgments}
The authors  wish  to  thank N. Chilton, J. Bartholomew, R. Ahlefeldt, M. Sellars, J. and  for helpful discussions. The proposal numbers for the data presented in this manuscript are \#20241028 (EIGER, SINQ~\cite{stuhr2017thermal}), \#19994 (Taipan, ANSTO~\cite{danilkin2012taipan,rule2018recent}). This research is supported by A*STAR under Project No. C230917009, Q.InC Strategic Research and Translational Thrust; the MTC Young Investigator Research Grant (Award \# M24N8c0110); CQT++ Core Research Funding Grant (A*STAR). The authors would like to thank ACNS for the Taipan beamtime through proposal P19994. This work is based on experiments performed at the Swiss spallation neutron source SINQ, Paul Scherrer Institute, Villigen, Switzerland. Y. Z. and O. V. Y. acknowledge support by the Swiss National Science Foundation (grant 224624).
\end{acknowledgments}

\bibliographystyle{unsrt}
\bibliography{ref12.bib}
\end{document}